# Simulations demonstrate a simple network to be sufficient to control branch point selection, smooth muscle and vasculature formation during lung branching morphogenesis

Géraldine Cellière[1], Denis Menshykau[1] and Dagmar Iber[1,2,*]

[1]Department for Biosystems Science and Engineering, ETH Zurich, Mattenstrasse 26, 4058 Basel, Switzerland
[2]Swiss Institute of Bioinformatics (SIB), Mattenstrasse 26, 4058 Basel, Switzerland

*Author for correspondence (dagmar.iber@bsse.ethz.ch)

## Summary

Proper lung functioning requires not only a correct structure of the conducting airway tree, but also the simultaneous development of smooth muscles and vasculature. Lung branching morphogenesis is strongly stereotyped and involves the recursive use of only three modes of branching. We have previously shown that the experimentally described interactions between Fibroblast growth factor (FGF)10, Sonic hedgehog (SHH) and Patched (Ptc) can give rise to a Turing mechanism that not only reproduces the experimentally observed wildtype branching pattern but also, in part counterintuitive, patterns in mutant mice. Here we show that, even though many proteins affect smooth muscle formation and the expression of *Vegfa*, an inducer of blood vessel formation, it is sufficient to add FGF9 to the FGF10/ SHH/Ptc module to successfully predict simultaneously the emergence of smooth muscles in the clefts between growing lung buds, and *Vegfa* expression in the distal sub-epithelial mesenchyme. Our model reproduces the phenotype of both wildtype and relevant mutant mice, as well as the results of most culture conditions described in the literature.

## Introduction

The main function of the lung is to enable efficient gas exchange. To this end a complex organ has evolved that, besides the airways, encompasses several interacting structures: the vasculature, the lymph system, the nerves and smooth muscles. The vasculature develops at the same time as the airways in a tightly coordinated process. Malformations of pulmonary blood vessels result in severe diseases like alveolar capillary dysplasia or bronchopulmonary dysplasia (Gao and Raj, 2010; deMello, 2004). The dense capillary plexus surrounding the distal epithelium forms and reorganizes as the lung tip grows out (Gebb and Shannon, 2000; Parera et al., 2005; Schachtner et al., 2000). VEGFA is a key player in this process, as alterations of VEGFA levels severely affect vascularization (Healy et al., 2000; Zhao et al., 2005; Akeson et al., 2003; Ng et al., 2001). Between embryonic day (E)12.5 and E14.5 *Vegfa* is expressed in the lung mesenchyme and epithelium (Greenberg et al., 2002; Ng et al., 2001) and binds to its receptor Fetal liver kinase 1 (Flk1) at the surface of mesenchymal progenitors (Shalaby et al., 1995; Schachtner et al., 2000). VEGFA signaling through Flk-1 induces the differentiation and/or proliferation of endothelial cells, and thus gives rise to the early immature vascular network (Shalaby et al., 1995). White et al. showed that the temporal and spatial expression of *Vegfa* is regulated by SHH and FGF9 (White et al., 2007). Both proteins signal to the lung mesenchyme (rather than endothelial cells) and together are both necessary and sufficient to induce *Vegfa* expression. Inhibition of SHH signaling by cyclopamine treatment results in a strong decrease in *Vegfa* expression, that can only in part be rescued by addition of exogenous FGF9 (White et al., 2007). Similarly, the defective blood vessel formation in the *Fgf9*$^{-/-}$ mutant cannot be recovered by addition of exogenous SHH (White et al., 2007).

Airway smooth muscles (SM) surround the walls of airways in the lung where they regulate the airway diameter and ensure the matching of perfusion and ventilation within the lungs (Sparrow et al., 1999; Yoshida and Owens, 2005). During development, airway smooth muscles are essential for normal lung branching (Nakamura and McCray, 2000; Kim and Vu, 2006; Sparrow and Lamb, 2003; Zhou et al., 1996). After birth, excess of smooth muscles may be involved in asthma and other lung diseases as smooth muscles have the potential to increase airway responsiveness and constriction in response to external stimuli, such as dust, allergens, cold air or stress (Lazaar, 2002; Hershenson et al., 1997; Raghu et al., 1988; Lazaar and Panettieri, 2003). Proper development of smooth muscles is therefore important for lung development and functioning.

Several mutants have been described with defects in airway smooth muscle formation, including mice that lack FGF10 (Mailleux et al., 2005; Ramasamy et al., 2007), FGF9 (Yin et al., 2011; White et al., 2006; del Moral et al., 2006b; Colvin et al., 2001), SHH (Miller et al., 2004; Pepicelli et al., 1998; Weaver et al., 2003), BMP4 (Weaver et al., 1999; Weaver et al., 2003; Jeffery et al., 2005), WNT (Goss et al., 2011; Yin et al., 2011; Shu et al., 2005; Shu et al., 2002; DeLanghe et al., 2008; DeLanghe et al., 2005; Cohen et al., 2009), or their receptors (DeLanghe et al., 2006; van Tuyl and Post, 2000). FGF10 has been shown to direct the outgrowth of the lung bud and to induce expression of *Shh* (Park et al., 1998; Abler et al., 2009; Bellusci et al., 1997). SHH signaling represses *Fgf10* expression and induces differentiation of progenitor cells into smooth muscle cells (SM) (White et al., 2006). FGF9 enhances the expression of *Fgf10* (del Moral et al., 2006b) and blocks the formation of smooth muscle cells independently of SHH signaling (Yi et al., 2009). FGF10 and FGF9 achieve their different regulatory outcomes by signaling through distinct receptors (Zhang et al., 2006). The FGF10 receptor FGFR3b is expressed in the epithelium and FGF10-dependent signaling enhances *Shh* expression (Abler et al., 2009), while FGF9 signals mainly through FGFR1c and 2c in the mesenchyme and blocks smooth muscle differentiation and the expression of *Noggin*, an antagonist of BMP signaling (Yin et al., 2011). Overexpression of *Xenopus Noggin* or *Gremlin*, a similar BMP antagonist, results in the proximalization of the distal lung tip and ectopic smooth muscles in distal areas (Lu et al., 2001; Weaver et al., 1999). The BMP receptor ALK3 is located in the epithelium, and BMP signaling primarily regulates the proliferation, survival and morphogenetic behavior of distal lung epithelial cells (Eblaghie et al., 2006). All FGFs enhance the expression of *Bmp4* in the distal lung epithelium (Hyatt et al., 2002) while BMP-dependent signaling can inhibit FGF-dependent signaling, possibly by enhancing expression of the FGF antagonist *Sprouty* (Hyatt et al., 2004). A number of other factors have been implicated in the control of smooth muscle formation. Smooth muscle differentiation is triggered by cell progenitor interactions with the basal membrane composed of laminin and fibronectin between the epithelial and mesenchymal cells (Yang et al., 1998; Yang et al., 1999; Zhang et al., 1999; DeLanghe et al., 2005). WNT signaling may impact smooth muscle development through the regulation of fibronectin deposition (DeLanghe et al., 2005); WNT2 has also been shown to regulate the expression of *Fgf10*, *Fgfr1c* and *2c* and *Wnt7b* (Goss et al., 2011; Yin et al., 2011). Furthermore, retinoic acid upregulates *Shh* expression and downregulates *Fgf10* expression (Cardoso et al., 1995; Kim and Vu, 2006; Malpel et al., 2000).

Intriguingly, in the developing lung, smooth muscles appear progressively around the proximal epithelium in between newly out-growing branches (Yi et al., 2009), while *Vegfa* and the vasculature appear in the distal part of the lung (White et al., 2007; Parera et al., 2005; Schachtner et al., 2000). Experiments suggest that FGF10 determines the points at which new buds grow out since FGF10 can direct the outgrowth of lung buds towards a source of FGF10 (Park et al., 1998; Bellusci et al., 1997). A combination of genetic experiments further suggests that SHH induces progenitor cells to differentiate into smooth muscle cells and distal mesenchyme to express *Vegfa* (White et al., 2006; White et al., 2007). We have recently developed a mathematical model to explain the self-organized emergence of FGF10 and SHH signaling spots in the developing lung bud (Menshykau et al., 2012). In this previous model, the equations for FGF10 (F), SHH (S), and its receptor Ptc (R) read:

$$
\begin{aligned}
[\dot{S}] &= \underbrace{\bar{D}_S \bar{\Delta}[S]}_{\text{diffusion}} + \underbrace{\bar{\rho}_S \frac{[\text{F10}]^n}{[\text{F10}]^n + \bar{K}_S^n}}_{\text{production}} - \underbrace{\bar{\delta}_C \bar{\Gamma}[R]^2[S]}_{\text{complex formation}} - \underbrace{\bar{\delta}_S[S]}_{\text{degradation}} \\
[\dot{R}] &= \underbrace{\bar{D}_R \bar{\Delta}[R]}_{\text{diffusion}} + \underbrace{\bar{\rho}_R + \bar{v}\bar{\Gamma}[R]^2[S]}_{\text{production}} - \underbrace{2\bar{\delta}_C \bar{\Gamma}[R]^2[S]}_{\text{complex formation}} - \underbrace{\bar{\delta}_R[R]}_{\text{degradation}} \\
[\dot{F10}] &= \underbrace{\bar{D}_{\text{F10}} \bar{\Delta}[\text{F10}]}_{\text{diffusion}} + \underbrace{\bar{\rho}_{\text{F10}} \frac{\bar{K}_{\text{F10}}^n}{(\bar{\Gamma}[R]^2[S])^n + \bar{K}_{\text{F10}}^n}}_{\text{production}} - \underbrace{\bar{\delta}_{\text{F10}}[\text{F10}]}_{\text{degradation}}.
\end{aligned}
\quad (1)
$$

Here we used $\dot{X} = \frac{\partial X}{\partial t}$ as short-hand notation for the time derivative. The model has been discussed in detail before (Menshykau et al., 2012). In brief, FGF10 enhances the expression of *Shh* $\left( \bar{\rho}_S \frac{[\text{F10}]^n}{[\text{F10}]^n + \bar{K}_S^n} \right)$ which, in turn, when bound to its receptor, reduces the expression of *Fgf10* $\left( \bar{\rho}_{\text{F10}} \frac{\bar{K}_{\text{F10}}^n}{(\bar{\Gamma}[R]^2[S])^n + \bar{K}_{\text{F10}}^n} \right)$. SHH-Receptor-binding reduces the concentration of free SHH $(-\bar{\delta}_C \bar{\Gamma}[R]^2[S])$ but enhances the receptor concentration by enhancing its expression $\left( \bar{v}\bar{\Gamma}[R]^2[S] \right)$. The receptor is expressed also at a constitutive rate $\rho R$, and all proteins linearly decay at rate $\bar{\delta}_X X$. FGFs and SHH can diffuse rapidly (Kicheva et al., 2007; Yu et al., 2009; Ries et al., 2009) and we write $\bar{D}_{\text{F10}}$ and $\bar{D}_S$ for the diffusion coefficients. Ptc receptors are membrane proteins and thus diffuse with a much reduced diffusion coefficient $\bar{D}_R \ll \bar{D}_S, \bar{D}_{\text{F10}}$ (Kumar et al., 2010; Hebert et al., 2005). In the tissue, diffusion of receptors is mainly restricted to the surface of single cells. A small, non-zero value for the receptor diffusion coefficient is nonetheless warranted, as previously discussed in detail. We write $\overline{D\Delta}[\cdot]$ for the diffusion fluxes where $\bar{\Delta}$ denotes the Laplacian operator in Cartesian coordinates, and $[\cdot]$ concentration. The characteristic length of gradients depends both on the speed of diffusion and the rate of morphogen removal.

The model showed that the biochemical interactions between FGF10, SHH, and its receptor Patched (Ptc), as graphically summarized in Fig. 1A, are sufficient to explain the emergence of FGF10 patterns consistent with the two main modes of branching in the developing lung: lateral branching and bifurcations. Fig. 2 presents an example of the FGF10 localization that would correspond to the lateral branching mode on a growing lung bud. As a note aside, the branching modes described by Metzger et al., 2008, i.e. domain branching and planar versus orthogonal bifurcations, are intrinsically 3D patterning events that cannot be addressed in our current simulations on a 2D slice of the developing lung (Fig. 1B).

In the following, we will integrate the regulatory interactions that control smooth muscle formation and VEGF-A expression into the previous model for FGF10 and SHH patterning in the lung bud and show that the model is consistent with all available experimental information to which the model applies, in particular mutants that affect smooth muscle formation and VEGF-A expression pattern.

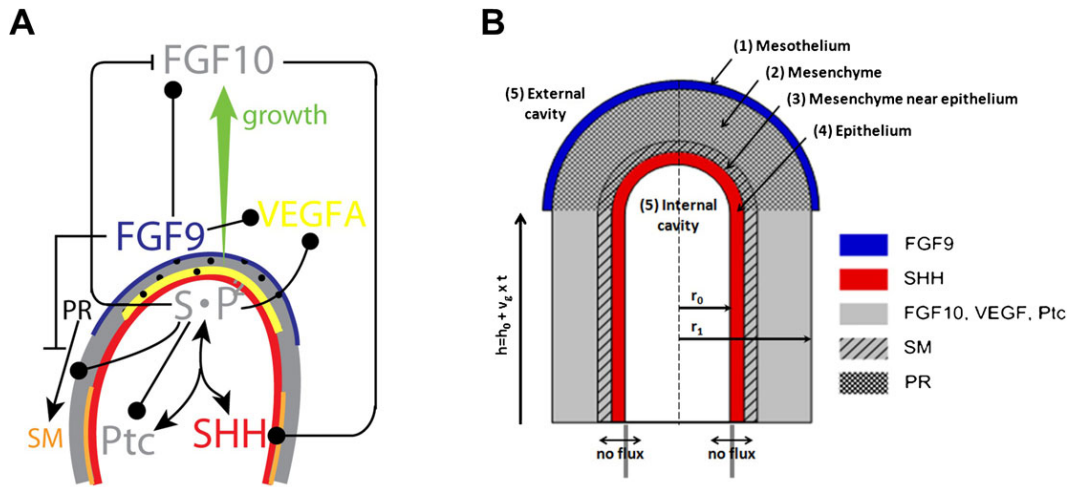

**Fig. 1. A graphical summary of the modelled interactions that control smooth muscle formation and *Vegfa* expression during lung bud morphogenesis.**
(**A**) FGF10 is transcribed at high levels in the distal mesenchyme (grey), and experiments suggest that FGF10 promotes both the proliferation of the endoderm and its outward movement (green arrow). FGF10 stimulates the expression of *Shh* in the epithelium (red). SHH reversibly binds its receptor, Ptc, which is expressed in the mesenchyme (grey). SHH-Ptc binding results in the repression of *Fgf10* expression, the upregulation of Ptc, and induces smooth muscle formation (SM) from the progenitor cells (PR). *Fgf9* is expressed in the mesothelium (blue) and, at later stages, also in the distal epithelium (red). It prevents the differentiation of progenitor cells into smooth muscle cells, and stimulates FGF10 production in the mesenchyme. *Vegfa* expression in the mesenchyme requires both FGF9 and Shh signaling through Ptc. (**B**) The idealized computational domain comprises a 2D cross section along the cylinder axis of symmetry. The mesothelium, the mesenchyme and the epithelium are shown in blue, grey and red, respectively. We distinguish two subdomains in the mesenchyme: near to the epithelium (hatched grey) laminin in the extracellular matrix allows progenitors to differentiate into smooth muscles, but this is not the case in the rest of the mesenchyme (light grey). SHH, FGF10, and FGF9 (but not Ptc, the progenitors and smooth muscles) can diffuse freely ($D_{ext}$) in the external and internal cavities (5). The legend on the figure indicates the species production sites. The time-dependent height of the cylinder is $h(t){=}h_0{+}v_g{\times}t$.

## Results

### An integrated model for lung branching, smooth muscle formation and vasculature development

We seek to expand our model for branch point selection to simulate also the coordinated emergence of the vasculature and smooth muscles. The signaling proteins that have been implicated in the control of smooth muscle differentiation form a complex network with many in part incoherent and indirect regulatory interactions. Computational models can help to disentangle the relative contributions and dependencies. In addition to the factors already incorporated, many other signaling proteins have been implicated in the control of the smooth muscle and blood vessel differentiation process, most importantly FGF9 which blocks smooth muscle formation, induces *Vegfa* and enhances FGF10 production (Yi et al., 2009; White et al., 2006; del Moral et al., 2006b). The effects of the other reported signaling proteins appear, however, to be mainly indirect by affecting the expression or activity of FGF10, SHH, Ptc or FGF9, i.e. retinoic acid upregulates *Shh* expression and downregulates *Fgf10* expression (Cardoso et al., 1995; Kim and Vu, 2006) and WNT regulates the expression of *Fgf10* and *Fgfr1c/2c* (Goss et al., 2011; Yin et al., 2011). Even though BMP signaling affects lung development and smooth muscle formation, only indirect effects via FGF signaling are well documented, and we therefore do not consider WNT, retinoic acid or BMP signaling in this parsimonious model. Finally, it has been suggested that WNT also regulates fibronectin deposition (DeLanghe et al., 2005). Laminin and fibronectin are two proteins of the extracellular matrix that form between the epithelial and mesenchymal cells. Evidence suggests that smooth muscle differentiation is triggered by their spreading on this basal membrane (Yang et al., 1998;

Yang et al., 1999; Zhang et al., 1999; DeLanghe et al., 2005). To reflect this requirement in our new model, we restricted the formation of smooth muscles to a thin layer of mesenchyme, adjacent to the epithelium.

Blood vessel formation is a partly stochastic process (Jain, 2003) that is not possible to directly describe in our model. We therefore chose to focus on VEGFA, the main inducer of endothelial cell differentiation (Jain, 2003; White et al., 2007). SHH and FGF9 were shown to directly control mesenchymal *Vegfa* expression (White et al., 2007). Epithelial VEGFA, on the other hand, does not seem to be influenced by SHH and FGF9 (White et al., 2007). We therefore integrated only the regulation of mesenchymal (but not epithelial) VEGFA in the model. VEGFA not only controls blood vessel formation but is also part of the epithelial-mesenchymal interactions regulating lung branching morphogenesis (Del Moral et al., 2006a). Stimulation of VEGFA signaling by exogenous VEGFA stimulates branching. Conversely, inhibition of VEGFA by antisense oligodeoxynucleotide to the Flk-1 receptor results in decreased epithelial branching (Del Moral et al., 2006a). The effects of VEGFA signaling on branching are likely mediated by increased cell proliferation, upregulation of BMP4 and downregulation of Sprouty2 and 4 in the epithelium (Del Moral et al., 2006a). Still Shh and Fgf10 expression were found unchanged in the presence of enhanced VEGFA signaling. As we consider only direct regulatory interactions, we omitted VEGFA's downstream functions.

To reflect their regulatory interactions, we add four differential equations to the core model for branch point selection (equation set Eq. 1) that describe FGF9 (denoted F9), VEGF-A (V), progenitor (P), and smooth muscle (M) cell population dynamics, and we modify our previous equations for FGF10 (F10), SHH (S), and Ptc receptor (R) accordingly. As previously, proteins (F10, S, F9, V) can diffuse freely in the tissue and the cavities, whereas cells (M and P) and receptor proteins (R) are restricted to the tissue. FGF9 is produced at a constant rate in the mesothelium or in the epithelium, and diffuses to the mesenchyme, where it is degraded by its cognate receptor (Orr-Urtreger et al., 1993; Zhang et al., 2006). We then write for the FGF9 dynamic:

$$[\dot{F9}] = \underbrace{\overline{D}_{F9}\overline{\Delta}[F9]}_{\text{diffusion}} + \underbrace{\overline{\rho}_{F9}}_{\text{production}} - \underbrace{\overline{\delta}_{F9}[F9]}_{\text{degradation}} \qquad (2)$$

Deviating from our previous model, FGF10 (F10) production in the mesenchyme is not only repressed by SHH (S) but also enhanced by FGF9 (F9). We note that FGF9 is, however, not necessary for *Fgf10* expression. We therefore reformulate Eq. 1 and write:

$$[\dot{F10}] = \underbrace{\overline{D}_{F10}\overline{\Delta}[F10]}_{\text{diffusion}} + \left[ \underbrace{\overline{\rho}_{F10} \quad \frac{\overline{K}_{F10}^n}{\overline{K}_{F10}^n + (\overline{\Gamma}[R]^2[S])^n}}_{\text{production inhibited by } [R]^2[S]} \right.$$

$$\left. \times \left( \underbrace{1}_{\text{constitutive}} + \underbrace{\overline{\kappa}_{F9} \frac{[F9]^2}{[F9]^2 + \overline{K}_{F10F9}^2}}_{\text{upregulation by F9}} \right) \right] - \underbrace{\overline{\delta}_{F10}[F10]}_{\text{degradation}} \qquad (3)$$

Moreover, we now model the spatio-temporal distribution of two cell populations, smooth muscles cells (M) and progenitors (P). Although smooth muscles and progenitors are (large) cells rather than (small) proteins, we model their density with continuous reaction-diffusion equations, without simulating cells individually. Experimentally, smooth muscles are usually visualized by antibodies against α-smooth muscle actin. Thus, we could also consider the M quantity as a concentration of α-smooth muscles actin. It should be noted that progenitor and smooth muscle cells are allowed to diffuse with a very small diffusion coefficient DC (orders of magnitude lower than for the receptor Ptc) to improve computational stability and accuracy; the exact value of DC has no impact on the distribution of proteins and may reflect random cell movement. Progenitor cells (P) are produced at a constant rate in the distal mesenchyme (at the tip). Progenitor cells can differentiate into smooth muscle cells (M) if two conditions are fulfilled: SHH is present and FGF9 is absent. Accordingly we write for the rate of differentiation:

$\eta[P]\left(\frac{\overline{\Gamma}[R]^2[S]}{\overline{\Gamma}[R]^2[S] + \overline{K}_M}\right)\left(\frac{\overline{K}_{F9}^m}{\overline{K}_{F9}^m + [F9]^m}\right)$. Progenitors can also die or differentiate into other cell types, i.e. fibroblasts, and we thus include a general loss term $\overline{\delta}_P[P]$.

$$[\dot{P}] = \underbrace{\overline{\rho}_P}_{\text{production}} - \underbrace{\overline{\delta}_P[P]}_{\text{loss}} - \underbrace{\eta[P]\left(\frac{\overline{\Gamma}[R]^2[S]}{\overline{\Gamma}[R]^2[S] + \overline{K}_M}\right)\left(\frac{\overline{K}_{F9}^m}{\overline{K}_{F9}^m + [F9]^m}\right)}_{\text{differentiation}} \qquad (4)$$

$$[\dot{M}] = \underbrace{\eta[P]\left(\frac{\overline{\Gamma}[R]^2[S]}{\overline{\Gamma}[R]^2[S] + \overline{K}_M}\right)\left(\frac{\overline{K}_{F9}^m}{\overline{K}_{F9}^m + [F9]^m}\right)}_{\text{differentiation}} \qquad (5)$$

SHH and FGF9 together are necessary for VEGF expression in the mesenchyme. Furthermore, the loss of one of the inducer can at most partly be rescued by addition of the other inducer. We then write for the VEGFA dynamic:

$$[\dot{V}] = \underbrace{\overline{D}_V\overline{\Delta}[V]}_{\text{diffusion}} + \left( \underbrace{\overline{\rho}_{V1}}_{\text{constitutive}} + \underbrace{\frac{[F9]^n}{[F9]^n + \overline{K}_{V1}^n}}_{\text{upregulation by F9}} \right)$$

$$\times \left( \underbrace{\overline{\rho}_{V2}}_{\text{constitutive}} + \underbrace{\overline{\rho}_{V3}\frac{(\overline{\Gamma}[R]^2[S])^n}{(\overline{\Gamma}[R]^2[S])^n + \overline{K}_{V2}^n}}_{\text{upregulation by SR}^2} \right) - \underbrace{\overline{\delta}_V[V]}_{\text{degradation}} \qquad (6)$$

Equations are non-dimensionalized in the same way as described by Menshykau et al. (Menshykau et al., 2012) (for details, see supplementary material Eq. S1). The non-dimensionalized system of equation no longer depends on absolute values (rate coefficients, diffusion coefficients and concentrations) but only on relative values. The values of all Hill coefficients ($m$, $n$) were set to two to account for possible cooperative effects; the model gives similar results with other values of the Hill coefficients (supplementary material Fig. S1). With the new model, both branching modes can still be obtained; however, upregulation of FGF10 by FGF9 makes lateral branching more favorable (supplementary material Fig. S2).

The proposed non-dimensional model has 31 parameters. 14 of these are part of the previously published model for lung branching morphogenesis. Four of these relate to the lung tip geometry and are based on experimental data. Sensitivity analysis for the core model shows that the value of most dimensionless parameters can be changed by 20–30% without qualitative change in the observed pattern. Furthermore, we showed that the core module is robust to parameter variation: parameter values were assumed to be given by the formula $k = k_0(1 + \xi(x,y))$, where $\xi(x,y)$ is normally distributed random function with a mean value of zero and half width $\theta$. The domain branching mode remains stable as long as standard deviations of the random variables $\theta$ do not exceed 0.2–0.3 of the reference value (Menshykau et al., 2012). The 17 new parameters were chosen, where possible, in line with the core parameter values, i.e. the diffusion constants for FGF9 and VEGF are the same as for FGF10. The degradation rates were then adjusted to obtain the experimentally observed diffusion lengths. The choice of the Hill constants relative to the production rates and of the Hill coefficients strongly determine the read-out pattern and their impact is explored in detail in supplementary material Figs S1–S3, as is discussed. It should be noted that we are not basing any conclusions on the particular choice of parameters but rather conclude that a parameter set can be found within the physiological range that enables us to reproduce all relevant published wildtype (WT) and mutant phenotypes.

The model is solved on a 2D slice of a lung-bud-shaped domain (Fig. 1B). Freely diffusible proteins FGF10, FGF9, SHH and VEGF-A are allowed to diffuse within epithelium and mesenchyme and into the cavities. The simulations start with no species present. To incorporate the effects of growth at the tip of the domain, the computational domain is elongated such that the height of the stalk is time-dependent, i.e. $h(t) = h_0 + v_g \times t$. The validity of this approach is discussed in detail by Menshykau et al. (Menshykau et al., 2012). Since new "matter" is added only at the front between tip and stalk to account for the local growth at the tip, protein concentrations are not diluted in this process and stay at the same absolute position. The growth rate $v_g$ was set as previously discussed by Menshykau (Menshykau et al., 2012). In brief, $v_g$ used to model the lateral branching mode in our model is around 14 $\mu$m.h$^{-1}$ and gives rise to two new branches per day with branches separated by approximately 150–200 $\mu$m; this is well in agreement with experimental observations (Metzger et al., 2008; Bellusci et al., 1997). To simulate the bifurcation mode of branching, we use a growth rate of 3.6 $\mu$m.h$^{-1}$, which is close to the growth speed estimated from the data of Metzger et al. (Metzger et al., 2008).

### FGF9, FGF10 and SHH are sufficient to coordinate SM emergence and *Vegfa* expression with branch patterning

With the model extensions, the simulations still show the spontaneous emergence of the spotted FGF10 pattern described by Menshykau et al. (Menshykau et al., 2012). The addition of the FGF9-FGF10 interaction results in an even more realistic pattern with FGF10 being more concentrated at the tip compared to more proximal areas (Fig. 2) (Bellusci et al., 1997). It also slightly promotes lateral branching over bifurcation (supplementary material Fig. S2).

The expanded model not only reproduces the FGF10 pattern but also yields realistic patterns of *Vegfa* and smooth muscle emergence (Fig. 3). Unlike other organs where *Vegfa* is expressed by the epithelium, lungs at E12.5–14.5 display mainly mesenchymal *Vegfa* expression (Greenberg et al., 2002; Ng et al., 2001). Induction by FGF9 and SHH signaling leads *Vegfa* to be upregulated distally, mainly in the sub-epithelial mesenchyme (White et al., 2007). We observe a similar expression pattern in our simulations. FGF9, produced in the distally localized mesothelium, accounts for the distal-proximal expression gradient of *Vegfa*, while SHH signaling via Ptc naturally enhances *Vegfa* in sub-epithelial mesenchyme (Fig. 2, Fig. 3A).

FGF9 and SHH are also the key proteins controlling smooth muscle differentiation. Experiments show that SM do not form in the tip but appear only behind the tip as the bud is growing out and this has been accounted to *Fgf9* expression in the mesothelium (Yi et al., 2009). We observe a similar behavior in our simulations. As in experiments, smooth muscles appear between FGF10 spots that attract newly outgrowing branches, except at the tip, where FGF9 inhibits the differentiation process (Fig. 3B). In summary, the alternating FGF10 and SHH signaling spots define the outgrowing (future distal) or non-outgrowing (future proximal) epithelium. The FGF9 gradient determines proximal versus distal fate once the bud has started to grow. As new buds always grow towards the FGF9 expressing mesothelium, FGF9 might act as a master regulator of all proteins and cell types that show a distal-proximal segregation pattern, i.e. not only *Vegfa* and smooth muscles, but possibly also Bmp4 (Weaver et al., 2003; Bellusci et al., 1996), Wnt (Yin et al., 2008; Bellusci et al., 1996), and TGF$\beta$ (Chen et al., 2007).

We note that the observed distribution and expression pattern correspond overall rather well to experimental observations, but since quantitative data on protein concentration distributions are not available we have to restrict ourselves to a qualitative discussion of distribution patterns. We therefore deliberately left out scale bars for better readability.

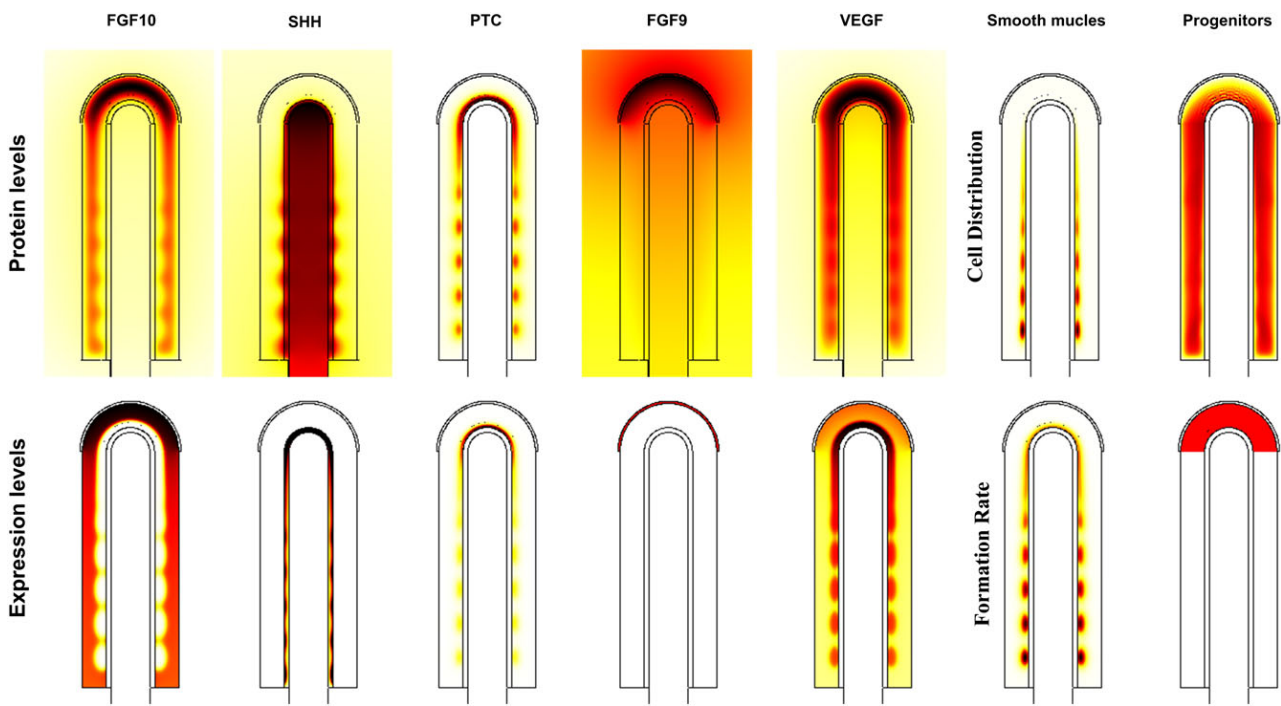

**Fig. 2. Protein distribution and expression pattern of all species.** Protein levels refer to the species concentrations while expression patterns refer to the production term of the equations. Scales (white-low, red-intermediate, black-high) are relative. Parameters are as in Table 1.

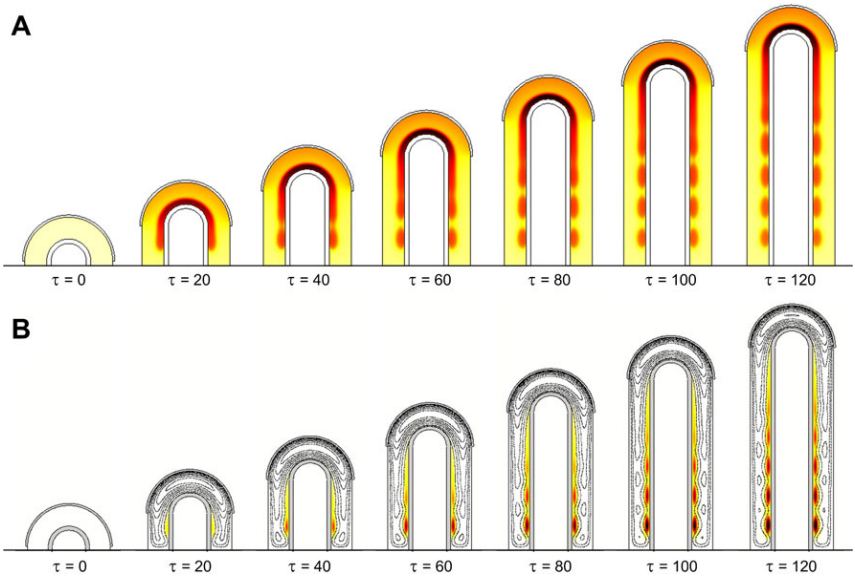

At early stages (E10.5–E11.5) the mesothelium covers the entire outgrowing lung domain. At that time no smooth muscles differentiate from the distal progenitor pool (Mailleux et al., 2005), but early markers of the vasculature can already be detected (Schachtner et al., 2000). As the secondary branches grow out, the mesothelium is pushed distally, allowing smooth muscles to emerge. To model this, we run simulations with different fractions of mesenchyme covered by mesothelium. As our in silico bud is growing, an increasing number of SM spots emerge behind the tip of the lung bud (supplementary material Fig. S3). The size of the mesenchyme covered by mesothelium has a major impact on smooth muscle formation: SM spots emerge earlier and more SM form over time in our simulations if the fraction of the stalk that is free of the FGF9-expressing mesothelium is increased (supplementary material Fig. S3A–C). It is difficult to establish from the published experimental data how far the length or the fraction of a mesothelium-free stalk may be conserved as the bud is growing out. We, however, obtain similar results in both cases (supplementary material Fig. S3). In our simulations, the distance of smooth muscle forming regions from the tip depends on the strength of SM inhibition by FGF9, which is a function of the smooth muscle progenitor sensitivity to the FGF9 concentration and the local FGF9 concentration. Sensitivity of progenitors to FGF9 is controlled by the Hill constant and the Hill coefficient. The local FGF9 concentration is determined by FGF9 expression and degradation rates, FGF9 diffusivity and the size of the FGF9 expressing domain. The overall impact of changes in the Hill coefficient $m$ are small, as long as $m$ is two or larger, and it affects only the concentration of smooth muscle cells and the speed with which they emerge, but not their positioning (supplementary material Fig. S1). Unlike smooth muscles, *Vegfa* expression is stimulated by FGF9 and therefore adopts the opposite dynamics. It is present in both the proximal and distal parts of the mesenchyme when the mesothelium covers the primary bud entirely. As the mesothelium is pushed by the secondary buds, VEGFA levels decrease (White et al., 2007). The early blood vessels that have started to form can then mature under the control of other factors like Notch, Angiopoietins, Wnt, or TGFβ, PDGF (Jain, 2003; Gao and Raj, 2010).

## Epithelial and mesothelial FGF9 result in the same smooth muscle and *Vegfa* expression patterns

At the very beginning of lung develompment (E10.5), *Fgf9* is expressed in both the epithelium and the mesothelium (Colvin et al., 1999). At E12.5, FGF9 is only detected in the mesothelium (Colvin et al., 1999; del Moral et al., 2006b), but becomes expressed again in both tissues at E13.5 (del Moral et al., 2006b), to finally be observed only in the distal epithelium at E14.5 (del Moral et al., 2006b). As it is a major role of FGF9 to control smooth muscle formation and blood vessel establishment via VEGFA, we wondered whether the two different *Fgf9* expression sites might affect these two processes differently. We first compared the smooth muscle profiles along the proximal-distal axis when FGF9 was produced either in the mesothelium or in the distal epithelium (Fig. 4A). No data are available regarding the relative expression intensities of *Fgf9* in the two tissues, and we therefore chose the rate of FGF9 production in the epithelium $\rho_f^{epithel}$ so that the maximal smooth muscle intensity is identical in both cases. Under those conditions the profiles of smooth muscle intensity are almost identical (Fig. 4): spots of smooth muscles appear at the same place with the same intensity, but epithelial FGF9 is more effective than mesothelial FGF9 in blocking smooth muscle formation at places of future bud outgrowth. The extent to which SM intensity increases with distance to the tip does not depend on where the *Fgf9* expression site is located. Since FGF9 also enhances *Fgf10* expression, we also compared the FGF10 concentration pattern. As FGF9 is produced at a lower level in the epithelium than in the mesothelium ($\rho_f^{epithel} = 30$ and $\rho_f^{mesothel} = 100$), the FGF10 pattern is more preserved in the tip region when FGF9 is produced in the epithelium. This also explains why smooth muscles are more precisely located in that case.

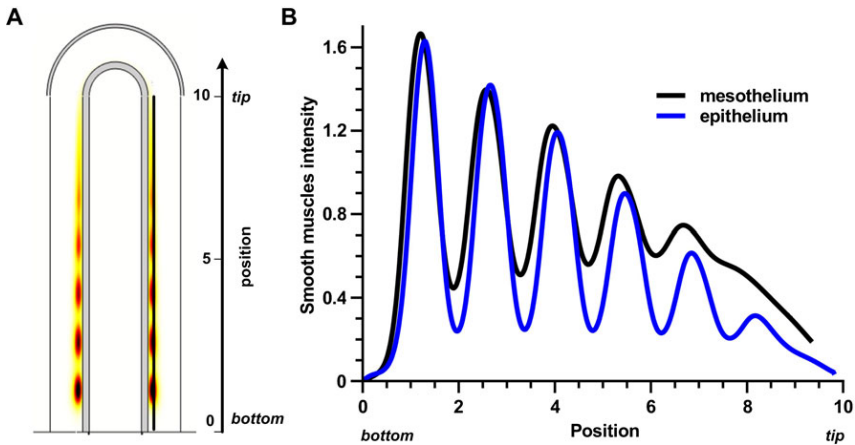

**Fig. 4. Similar smooth muscle profile when FGF9 is produced in the mesothelium or the epithelium.** (A) Distribution of smooth muscles at $\tau=120$ with FGF9 produced in the mesothelium. (B) Smooth muscles intensity at $\tau=120$ in the stalk. Position 0 is the bottom, position 10 the tip of the stalk. Production rates are adjusted for the two production sites ($\rho_I^{meso}=100$, and $\rho_I^{epithel}=30$) in order to have the same maximal smooth muscles intensity. All parameters are as in Table 1 unless otherwise stated.

FGF9 upregulates *Vegfa* expression at the tip, and SHH signaling restricts it near the epithelium. Both epithelial and mesenchymal FGF9 create the proximal-distal gradient required for correct *Vegfa* expression. The precise shape of the gradient only mildly affects *Vegfa* distribution (data not shown). We therefore conclude that the location of *Fgf9* expression has only little impact on SM differentiation, *Fgf10* and *Vegfa* expression, except that a lower overall FGF9 production rate is required in the epithelium to achieve the same effect.

Yin et al. suggest that the two sources of FGF9 have unique and specific regulatory functions (Yin et al., 2011). According to them, mesothelial FGF9 mainly regulates mesenchymal proliferation, whereas epithelial FGF9 is thought to influence branching in an autocrine fashion (Yin et al., 2011; del Moral et al., 2006b). We indeed observe that mesothelial FGF9 diffuses rather widely into the mesenchyme (because of a broad expression site), whereas epithelial FGF9 is more restricted near the epithelium (supplementary material Fig. S4). Thus the mechanism proposed by Yin et al. is possible.

### Influence of neighboring lung buds and the growth mode on smooth muscle emergence and *Vegfa* distribution

When a primary bud grows out from the trachea, it is first fully enveloped by mesenchyme and a mesothelium layer. Subsequent branch generations grow out next to each other, so that one bud is not only surrounded by mesenchyme but may be influenced by secreted factors of adjacent buds (Fig. 5A). Instead of simulating an array of juxtaposed buds (which would be computationally costly) we captured the effect by imposing reflecting boundary conditions (no-flux) such that secreted factors that diffuse away will be reflected and thus available to the secreting lung bud (Fig. 5B). We find no impact by such no-flux boundary conditions and obtain the same pattern as before for all factors or cells, especially FGF10, smooth muscles and *Vegfa*. We therefore conclude that neighboring lung buds are unlikely to affect each other's development via shared secreted factors.

We next wondered whether the growth mode might affect SM emergence or *Vegfa* expression. Two different growth modes have been observed in the lung: uniform growth, where proliferation occur in the entire mesenchyme, and tip growth, where proliferating cells are mainly observed in distal areas (Okubo et al., 2005). We have so far only studied growth at the tip. To simulate uniform growth, the height of the bud increases as before, but the reaction-diffusion equations need to be modified to achieve conservation of mass (rather than of concentration) as described previously (Menshykau et al., 2012). In case of uniform growth, new FGF10 spots are no longer restricted to appear behind the tip but can emerge in any place within the domain (supplementary material Fig. S5) (Menshykau et al., 2012). In our simulations, SHH-Ptc signaling between the initial FGF10 spots induces SM formation. As a result there may already be some weak SM concentration by the time that a new FGF10 spot intercalates (supplementary material Fig. S5). Bead experiments demonstrate that an FGF10-bead implanted in a whole lung explant is able to attract distal epithelium but not proximal epithelium (Park et al., 1998), and it has been proposed that smooth muscles might be responsible for the non-responsiveness of proximal epithelium to FGF10 (Kim and Vu, 2006), possibly by acting as a diffusion barrier or through mechanical forces. We therefore predict that new branches can form in a uniform growth mode only if SM appear sufficiently slowly or sufficiently far from the tip, to be insufficiently dense to prevent the bud outgrowth when a new FGF10 spot appears in more proximal areas. Alternatively the phenotype of smooth muscles could have reversed (in the presence of high FGF10 concentrations), as is the case with vascular smooth muscles (Wang and Olson, 2004; Owens, 1995).

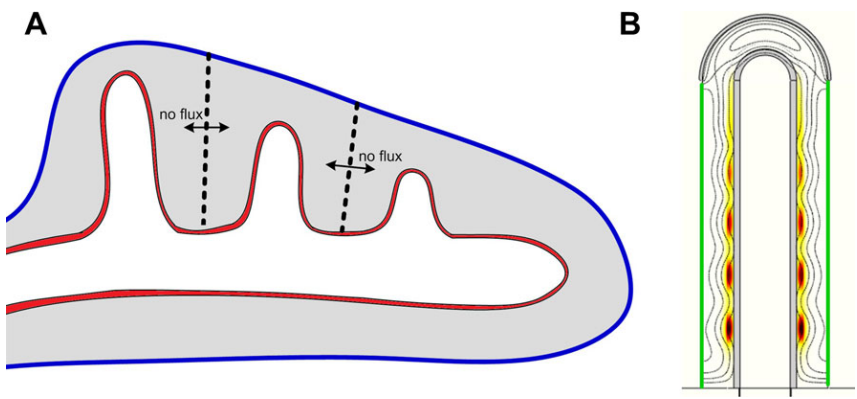

**Fig. 5. No-flux conditions modelling a bud surrounded by two other buds show coherent smooth muscles formation.** (A) Schematic representation of an early lung bud. Due to symmetry considerations, the lung bud in the middle can be modeled with no-flux boundary conditions. (B) Smooth muscles distribution (colormap white (low concentration), red (intermediate), black (high)) and FGF10 (black contours) at $\tau=120$ in case of no-flux conditions (in green) at the two exterior boundaries of the stalk. All parameters are as in Table 1.

This issue does not apply to *Vegfa*. Blood vessels form a dense plexus around both the proximal and distal epithelium. When a bud grows, either at the tip or more proximally in case of uniform growth, the blood vessels are able to remodel (Parera et al., 2005). With regard to blood vessels or *Vegfa* expression, a bud outgrowing in the stalk is therefore equivalent to that outgrowing at the tip.

### The model is consistent with observed mutant phenotypes

An important test for the suitability of a mathematical model is its consistency with a wide range of independent experimental observations. Lung branching morphogenesis has been studied intensively and a large body of experimental results exists to test the model with. These include a large number of in part counterintuitive mutant phenotypes of key signaling proteins in mice. The experimental papers typically report qualitative changes in local gene expression patterns as measured by *in situ* mRNA hybridization, quantitative changes in total mRNA production within the entire lung by real-time (RT)-PCR analysis, and changes in the detection of antibody staining against α-smooth muscle actin. We can compare the reported *in situ* hybridization data to changes in the localization of the production of the different proteins in our simulations, the RT-PCR data to total changes in protein production, and the antibody stainings to predicted changes in the localization and quantity of smooth muscles (Figs 6, 7). With regard to our model, we found

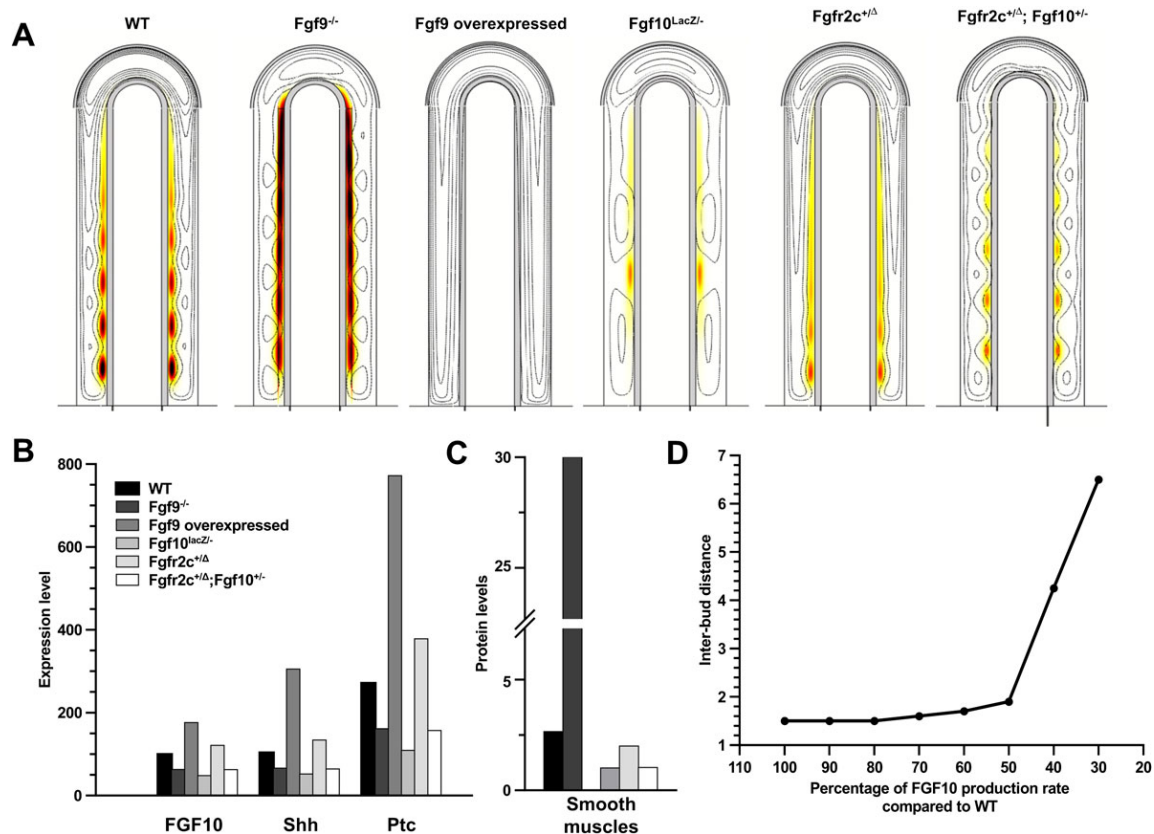

**Fig. 6. Mutants affecting smooth muscle formation can be reproduced.** (**A**) Distribution of smooth muscles (thermal colormap) and FGF10 (black contours) in different mutants. WT: smooth muscles appear between FGF10 spots. *FGF9⁻/⁻* mutant: setting FGF9 production rate to 0 leads to ectopic smooth muscles at the tip of the growing bud and more smooth muscles in proximal regions. Overexpression of FGF9 in the distal epithelium: FGF9 was produced in the mesothelium at rate $\rho_f$ (as in WT) and simultaneously in the epithelium at rate $3\times\rho_f$ (overexpression). This leads to a complete block of smooth muscle expression. Hypomorphic FGF10 mutant: when *Fgf10* expression is reduced to 45%, smooth muscle intensity is reduced. The distance between branching points is increased (see D). *Fgf2c⁺/ᐃ* mutant: Mesenchymal cells express both *Fgfr2c* (responding to FGF9, as in WT) and *Fgfr2b* (responding to FGF10, usually only in the epithelium). Instead of responding to the FGF9 concentration, mesenchymal cells now respond to the FGF9 plus FGF10 concentration. Smooth muscles are only slightly reduced (from 2.6 to 2) and patterning (FGF10 spots) could be observed only in proximal areas. *Fgfr2c⁺/ᐃ; Fgf10⁺/⁻* mutant: in order to compensate the effect of the mutation *Fgfr2c⁺/ᐃ*, one of the two alleles of *Fgf10* was deleted. Reducing FGF10 production rate two-fold, in addition to modifications for *Fgfr2c⁺/ᐃ*, restores the patterning of FGF10. (**B**) Expression levels (integral of the production term over the surface) of *Fgf10*, *Shh* and *Ptc* for all mutants. (**C**) SM concentration (integral of SM concentration over the surface) for all mutants. (**D**) Behavior of the allelic series for *Fgf10*. FGF10 production rate was stepwise decreased from 100% of WT to 30% of WT. The distance between two branching points is reported for each simulation. If FGF10 production rate is more than 50% of WT, the interbud distance is approximately constant. All simulations are shown at $\tau=120$. For simplicity, we kept for all mutants the same value for $v_g$ as in WT. All parameters are as in Table 1 unless otherwise stated.

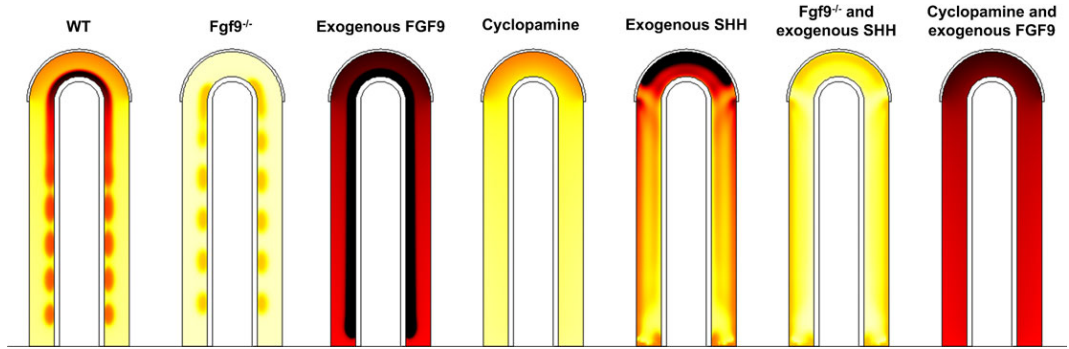

**Fig. 7. Mutants or culture conditions affecting *Vegfa* expression can be reproduced.** Pattern of *Vegfa* expression (thermal colormap, yellow (low concentration) and black (high concentration)) in WT and several mutants. WT: *Vegfa* is upregulated distally and in sub-epithelial mesenchyme. *Fgf9*[-/-] mutant: setting FGF9 production rate to 0 leads to a strong decrease in *Vegfa* expression. Exogenous FGF9: a constant concentration of FGF9 was added to the cavities. As a result, *Vegfa* is strongly increased in the mesenchyme. Cyclopamine treatment: Ptc was removed from the system to model inhibition of SHH signaling by cyclopamine. Compared to WT, no upregulation in sub-epithelial mesenchyme is observed. Exogenous SHH: A constant concentration of SHH in the cavities induces an enhancement of *Vegfa* expression at the tip. *Fgf9*[-/-] and exogenous SHH: exogenous SHH cannot rescue the *Fgf9*[-/-] phenotype. *Vegfa* levels stay low. Cyclopamine and exogenous FGF9: exogenous FGF9 partly rescues the cyclopamine treatment phenotype. All simulations are shown at $\tau$=120. For simplicity, we kept for all mutants the same value for $v_g$ as in WT. All parameters are as in Table 1 unless otherwise stated.

perturbations of FGF10, FGF9, and SHH signaling to test our model with.

- Hypomorphic FGF10: A transgenic mouse line that expresses *LacZ* under the control of the *Fgf10* regulatory sequences exhibits decreased *Fgf10* expression such that an allelic sequence, i.e. *Fgf10*[+/+] (WT), *Fgf10*[LacZ/+], *Fgf10*[+/−], *Fgf10*[LacZ/−], with decreasing *Fgf10* expression could be created (Mailleux et al., 2005; Ramasamy et al., 2007). All but the last line have normal lung phenotypes; the *Fgf10*[LacZ/−] mice have hypoplastic lungs with decreased epithelial branching and increased distance between branching points at E14.5. The extent of FGF10 expression has been measured only for the three mutant lines, but not for the WT. Moreover, the measured change in FGF expression will reflect both the direct changes due to the mutation as well as indirect effects, due to regulatory feedbacks. We therefore explored the effect of a lower *Fgf10* expression rate by decreasing stepwise the FGF10 production rate $p_F$ from 100% to 30%. Levels of *Fgf10*, *Shh* and *Ptc* expression, as well as levels of smooth muscle concentration, decrease progressively, and proportionally to the FGF10 production rate (supplementary material Fig. S6). The inter-bud (inter-spot) distance is constant, as the *Fgf10* expression rate is reduced from 100% to 50%, but it greatly rises below 50% (Fig. 6D). This is in perfect agreement with the experimental results that show that WT, *Fgf10*[LacZ/+], and *Fgf10*[+/−] have normal phenotypes and variable *Fgf10* expression levels, while *Fgf10*[LacZ/−] has an abnormal phenotype with an augmented distance between branching points. From the approximate change in the inter-bud distance observed experimentally we can deduce the corresponding change in $p_F$ and in smooth muscle intensity for the *Fgf10*[LacZ/−] mutant, compared to WT. In our model, smooth muscle intensity is reduced by 61% (Fig. 6C), which lies in the range of the experimental data (40% and 90% in the right and left secondary bronchi respectively).

In the *Fgf10*[+/−] and *Fgf10*[LacZ/−] mice, *Vegfa* expression was also assessed by real-time PCR (Ramasamy et al., 2007; Mailleux et al., 2005). Ramasamy et al. found a 28% decrease in *Vegfa* expression in *Fgf10*[LacZ/−] mice compared to the

*Fgf10*[+/−] mutant serving as a control at E14.5. Using again the interbud-distance in *Fgf10*[LacZ/−] lungs, we could deduce from Fig. 6C that Fgf10 expression in this mutant is around 40% of WT. From the data of Ramasamy et al., we can infer that Fgf10 expression in *Fgf10*[+/−] should be around 55% of WT. Thus we could simulate both mutants by setting the appropriate FGF10 production rates. Our model predicts a 6% decrease in *Vegfa* expression in *Fgf10*[LacZ/−] compared to *Fgf10*[+/−]. This is less than measured experimentally, but still in good agreement, when considering the many uncertainties in measurements.

- *Fgfr2c*[+/Δ]: FGF10 signals mainly via its FGFR2b receptor (Zhang et al., 2006; Orr-Urtreger et al., 1993), that is expressed in epithelial tissues, while FGF9 preferentially bind to FGFR2c, and FGFR3c, expressed in the mesenchyme (Zhang et al., 2006; Orr-Urtreger et al., 1993). FGFR2b and FGFR2c are two isoforms of the FGFR2 receptor, arising by alternative splicing. Deletion of the "c" exon in the *Fgfr2c*[+/Δ] line interferes with the normal splicing balance, so that mutant mice mesenchyme now expresses both *Fgfr2b* and *Fgfr2c* (DeLanghe et al., 2006). As both receptors have identical intracellular domains, mutant mesenchymal cells (including progenitors) sense both FGF9 and FGF10, without distinguishing between them. This creates a positive feedback loop for FGF10, as FGF10 now enhances its own expression. As a result, branching is severely perturbed, *Fgf10* expression is more widespread and slightly more intense in the mutant at E13.5. Even though both FGF10 and FGF9 inhibit smooth muscle formation in the mutant, airway smooth muscles still appear. The relative receptor levels in this mutant are not known and there are therefore several ways for us to implement the mutant. For simplicity we replaced FGF9 (*F9*) by FGF9+FGF10 (*F9*+*F10*) in Eq. 6. Simulations predict an increase in *Fgf10* expression (20%), which is no longer patterned in spots except in the very proximal part (Fig. 6A,B), as is also observed in experiments. Much as in experiments, the appearance of smooth muscles only slightly decreases in our simulations (from 2.65 to 2). The effect of FGF10 on SM formation is minor in the receptor mutant because the FGF10 concentration is much lower than the FGF9 concentration at the locations where SM form. Experiments further show that the emergence of SM can be blocked in the mutant (but not in WT) by treating E11.5 lung explants with FGF10 for 72h (DeLanghe et al., 2006). To simulate this experiment, we set a constant concentration of FGF10 in the lung cavities. This leads to a complete disruption of FGF10 patterning, both in WT and mutants. Much as in the experiments, we observe that smooth muscle intensity is similar in treated and untreated WT, but it is reduced in the FGF10-treated receptor mutant. The simulation results thus match very well all available experimental observations.

- $Fgfr2c^{+/\Delta}$; $Fgf10^{+/-}$: In order to rescue the phenotype of the $Fgfr2c^{+/\Delta}$ mice, these mutants were crossed with a heterozygous $Fgf10$ mouse line (DeLanghe et al., 2006). This partially rescued branching. We reproduce this mutant by lowering the FGF10 production rate to 50%. Simulations show a slightly reduced $Fgf10$ expression, with an almost normal FGF10 pattern (Fig. 6A), which indeed correspond to a rescue.

- $FGF9^{-/-}$: $FGF9^{-/-}$ mutant mice lungs are hypoplastic but still develop (Yi et al., 2009). At E12.5, smooth muscles are ectopically detected at the tip of growing buds, and α-smooth muscle actin staining is more intense in proximal areas compared to control lungs (Yi et al., 2009). In agreement with the experimental data, absence of FGF9 in our model leads to a strong (6-fold) increase of smooth muscle intensity (Fig. 6C) in the proximal areas and ectopic smooth muscle formation at the tip (Fig. 6A). The protein expression levels also match very well: $Fgf10$ expression is reduced by approximately 45% in this mutant (Yi et al., 2009), as is also the case in our simulation (Fig. 6B). The expression of both $Shh$ and $Ptc$ appears to be greatly reduced both experimentally (White et al., 2006) and in our model (reduction by 35% and 40% respectively) (Fig. 6B). $FGF9^{-/-}$ also has an abnormal vascular phenotype. $Vegfa$ is strongly restricted near the epithelium and no distal upregulation is observed (White et al., 2007). Our model simulating $Fgf9^{-/-}$ shows the same pattern as in vivo for $Vegfa$ production (expression). The discontinuous expression in the sub-epithelial mesenchyme is due to SHH signaling being confined between FGF10 spots. In vivo, the reduced $Vegfa$ expression then leads to an almost two-fold decrease in capillary coverage, as shown by PECAM in situ hybridization (White et al., 2007).

- FGF9 overexpression: $Fgf9$ has been overexpressed in the distal epithelium under the $SP$-$C$ promoter (White et al., 2006). No data are available on the extent of $Fgf9$ overexpression, but based on in situ hybridization data (White et al., 2006) we can infer that the amount of FGF9 production in the epithelium is at least 3-fold that in the mesothelium, and that $Fgf9$ is expressed in the entire epithelium. In our simulations we obtain a complete block of smooth muscle formation when $Fgf9$ is 3-fold overexpressed, as it is the case in the mutant mice at E14.5 (Fig. 6A–C). This is consistent with FGF9 inhibiting smooth muscle formation, even in presence of SHH (Weaver et al., 2003). Since the Turing pattern is lost when $Fgf9$ (and $Fgf10$) are strongly overexpressed, $Fgf10$ is expressed homogenously in the simulations (Fig. 6A, black contour lines), as also observed in the mutant mice (White et al., 2006). Our model also reproduces the experimentally observed increase in $Shh$ and $Ptc$ expression (Fig. 6B) (White et al., 2006). In a related experiment FGF9, has been added to the media of WT lung explants (del Moral et al., 2006b). The observations are similar to that of the $Fgf9$ overexpression mutant: smooth muscles are greatly reduced (but not absent), $Ptc$ is upregulated, branching is impaired. However, $Shh$ expression is found unchanged compared to control. We modeled this experiment by setting a constant concentration of FGF9 ($F9$=2, same order of magnitude as the concentration of FGF9 in the mesenchyme in the WT simulation) in the cavities. We observe that smooth muscles are indeed reduced (from 2 to 0.4). Deviating from experimental results, both $Ptc$ and $Shh$ are upregulated in our simulation, but to a lesser extent than in the overexpression mutant. Moreover, adding exogenous FGF9 in the medium of lung explants leads to a very strong expression of $Vegfa$ in the entire mesenchyme. Much as in the experimental data, our simulation shows a very strong and quasi-uniform VEGFA production.

- $Shh^{-/-}$: complete $Shh$ knock-outs have severe lung phenotypes as only two rudimentary sacs. Even at E18.5 no smooth muscles can be detected along those two sacs (Pepicelli et al., 1998; Miller et al., 2004). In our model, setting SHH production rate to zero disrupts the Turing pattern, and in the absence of their inducer, no smooth muscles appear (data not shown).

- Cyclopamine: Cyclopamine is an inhibitor of SHH signaling. At the concentration used in experiments, SHH signaling is completely blocked after 48 hours of treatment, as confirmed by absence of Ptc expression (White et al., 2006). Only $Vegfa$ (and not smooth muscle) distribution was studied under those culture conditions. Experimental data show that at E14.5 only a monolayer of $Vegfa$ expressing cells persists, adjacent to the epithelium. To model the cyclopamine treatment, we deleted Ptc from our system, thus preventing SHH signaling. Deprived of the upregulation by SHH, $Vegfa$ expression is found low and uniform in sub-mesothelial and sub-epithelial mesenchyme, in our simulation. This discrepancy between the experimental results and our model will be analyzed in the discussion.

- Cyclopamine and exogenous FGF9: To test if SHH and FGF9 could compensate for each other in $Vegfa$ induction, exogenous FGF9 was added to cyclopamine treated lung explants for 48 hours. FGF9 can partly rescue the cyclopamine phenotype, as $Vegfa$ is now detected in the entire mesenchyme, but at lower levels than with exogenous FGF9 alone. The simulations show similar results. VEGFA is widely express but not enhanced in sub-epithelial mesenchyme, consistent with the simulation results with cyclopamine treatment alone.

- Exogenous SHH: In response to exogenous SHH, $Vegfa$ expression is increased distally in sub-mesothelial mesenchyme. Simulations with a constant SHH concentration in the cavities reproduce this result. Exogenous SHH diffuses into the tissue and upregulates $Vegfa$ expression there. The smooth muscle phenotype resulting from the addition of exogenous SHH was not described.

- $Fgf9^{-/-}$ and exogenous SHH: Exogenous SHH was added for 24 hours to the medium of lung explants from $Fgf9^{-/-}$ mice, to test if SHH alone is sufficient to stimulate capillary formation. No data on $Vegfa$ expression are available for this experimental set-up but we can infer from the low vessel density, visualized by PECAM *in situ* hybridization, that the low *Vegfa* expression in *Fgf9^{−/−}* mutants cannot be rescued by exogenous SHH. This is also the case in our simulations, where exogenous SHH slightly modify the spatial pattern of *Vegfa* expression but not its level, in comparison to *Fgf9^{−/−}* lungs.

In conclusion, our model reproduces all experimental perturbations rather wall, with the exception of the cyclopamine experiment, where we cannot reproduce the restriction of *Vegfa* expression to a thin monolayer of cells.

## Discussion

During lung branching morphogenesis the airways develop together with the vasculature and smooth muscles. We have recently proposed a model that explains lung branch point selection with the biochemically established interactions between only three proteins FGF10, SHH, and Ptch (Menshykau et al., 2012). Here we have shown that a small extension of the model to include FGF9 enables us to recapitulate also smooth muscle formation and vasculogenesis. The model suggests that a highly integrated regulatory process coordinates the simultaneous emergence of the three essential structures: the airways, the blood vessels, and the smooth muscles.

We previously described lung branch site selection with only three coupled partial differential equations for SHH, Ptch, and FGF10. To extend the model to smooth muscle formation and VEGF expression we only needed to include one further coupled PDE for the growth factor FGF9; three further non-coupled PDEs were included as read-outs for progenitor cells and smooth muscle cells as well as for VEGF. Most of the parameter values have not yet been established in experiments. We therefore limited the number of free parameters by non-dimensionalizing the model. As a result we only have relative rather than absolute parameter values, i.e. relative diffusion coefficients, production and decay rates etc. This still leaves us with 31 free parameters in the extended model. Of these, only the geometrical parameters and those that define the diffusion length of components can be set directly based on experiments. We have previously shown for the core model that the patterns are robust to small variations in the 14 parameter values as may arise from molecular noise, but that the branching mode can change or the pattern can be completely lost in response to larger changes (Menshykau et al., 2012). The model extension introduced 17 new parameters and these needed to be set such that model results corresponded to data from available experimental perturbations, i.e. mutants and lung cultures. We had data from 13 independent experimental perturbations to adjust 17 new parameters that arose from the model extension. Each condition yielded expression patterns for several genes. It is thus not trivial to obtain a parameter set that is consistent with all available experimental information. We also explored the impact of the critical parameters extensively (supplementary material Figs S1–S3).

The newly introduced FGF9 also impacted on the core module. FGF9 enhances FGF10 expression in the distal lung bud and the predicted FGF10 expression pattern resembles FGF10 expression in the embryo even more closely than in the previous model. FGF10 marks newly outgrowing buds and thus determines the branching pattern. FGF9-dependent upregulation of distal FGF10 expression renders the lateral mode of branching more favorable (supplementary material Fig. S2). As levels of FGF9 expression decrease in time during embryonic development (Colvin et al., 1999; del Moral et al., 2006b), bifurcation modes of branching should become more favorable; experimental data indeed show that the domain branching mode is deployed in the beginning of lung branching to build a lung scaffold, while planar and orthogonal bifurcations are used at the later stages to fill surfaces and edges (Metzger et al., 2008).

Our model not only reproduces the WT phenotype, but also all reported mutants with abnormal smooth muscle or *Vegfa* expression phenotypes, and most of the *Vegfa* expression results from culture experiments. Among them, the FGF10 allelic series and the ectopic expression of the FGF10 receptor *Fgfr2b* in the mesenchyme result in effects that are not readily predictable with verbal reasoning. Thus the modest reduction of *Fgf10* expression in the hypomorphic mutants *Fgf10^{LacZ/+}* and *Fgf10^{+/−}* have no visible impact, while the more pronounced reduction of *Fgf10* expression in *Fgf10^{LacZ/−}* lead to an increased inter-bud distance. In the simulations, the distance between two branching points is increased only if the FGF10 production rate is lowered to less than 50%, which recapitulates very well the experimental observations. Reducing FGF10 levels also affects *Vegfa* expression, even if FGF10 is not a direct regulator of it. Our model predicts the non-trivial reduction in *Vegfa* expression, which is indeed observed *in vivo* in *Fgf10^{LacZ/+}* but to a greater extent (Ramasamy et al., 2007). The difference in the absolute predicted and observed reduction is likely the result of the accumulated uncertainties when estimating the extent to which model parameters need to be changed in the mutants based on incomplete or contradictory experimental measures. Another unexpected result concerns the FGF receptor mutant *Fgfr2c^{+/Δ}* that enables FGF10 signaling in the mesenchyme. Even if smooth muscles are now inhibited by both FGF9 and FGF10, they still form in the mutant. The simulations reproduce the result and provide an explanation in that FGF10-dependent signaling is weak relative to FGF9-dependent signaling in the SM forming spots. The fact that the model can reproduce also counterintuitive mutant phenotypes makes us confident that the model captures the key aspects of the regulatory processes that determine SM formation. However, our model is unable to explain one aspect of the *Vegfa* expression phenotype observed when lung explants are cultured in presence of cyclopamine, which inhibits SHH signaling. Instead of a monolayer of cells expressing *Vegfa* near the epithelium, as found experimentally, our model shows a more uniform *Vegfa* expression pattern. Indeed, in the simulations, in the absence of SHH signaling, only FGF9, which is rather widespread, induces *Vegfa*. White et al. interpret the very restricted *Vegfa* expression they observe by proposing that SHH signaling is necessary for *Vegfa* expression only in the sub-mesothelial mesenchyme. Yet it is unlikely that Shh produced in the epithelium acts only in the sub-mesothelial mesenchyme. On the contrary, well-documented effects of SHH signaling, like *Fgf10* inhibition, are clearly limited to sub-epithelial mesenchyme (Malpel et al., 2000; Bellusci et al., 1997). We therefore advance that *Vegfa* is regulated by more factors. Considering that *Vegfa* is expressed only in the layer just adjacent to epithelium in case of cyclopamine treatment, extracellular matrix components accumulating at the epithelial-mesenchymal border are possible candidates.

Mathematical models are powerful tools to rigorously test if the verbally formulated regulatory interactions proposed by experimentalists can indeed explain the observed phenotypes. Here we show that the previously identified regulators of smooth muscle formation are indeed sufficient to understand the WT phenotype and all 6 relevant mutant phenotypes. In contrast, the restriction of VEGF expression to the mesenchyme directly adjacent to the epithelium upon cyclopamine treatment cannot be explained with currently known regulatory interactions. Further experimental investigations are required. We also used our model to explore the impact of the different FGF9 expression sites. During lung development FGF9 is expressed sometimes in the mesothelium and sometimes in the epithelium. However, the model predicts little influence on Vegfa expression and smooth muscle formation as long as laminin restricts smooth muscle emergence to the epithelial-mesenchymal border. The role of the two independent production sites and the impact of their temporal use therefore remains an open question.

The geometry of the lung is complex and intrinsically 3-dimensional. We have previously shown that the behavior of the model in the 2D lung slice that we consider here is similar to the behavior in 3D (Menshykau et al., 2012). Secondary buds that arise from the primary branch by lateral branching are surrounded by mesothelium on two of their sides (ventrally and dorsally), while the two other sides face two neighboring buds (Metzger et al., 2008; Bellusci et al., 1997). Moreover, when buds grow out laterally, they push the mesothelium layer distally. Given this diversity, we tested several idealized geometries. Our model gives similar results both for permeable boundary conditions and no-flux conditions (modeling adjacent growing buds). We also allowed the bud to grow with different fixed fractions or lengths of mesenchyme covered by mesothelium. As expected, smooth muscles form only in the mesenchyme that is not covered by mesothelium (supplementary material Fig. S3), while *Vegfa* is upregulated in the mesothelium-covered mesenchyme. Simulations show that even though the extent of coverage of mesenchyme by FGF9-expressing mesothelium is a major regulator of smooth muscle formation (supplementary material Fig. S3) and *Vegfa* expression, the localization of FGF production (mesothelium versus distal epithelium) has little impact on the two processes. Yet, we note that inhibition by epithelial FGF9 is more efficient and a lower production rate $\rho_{fl}^{epithel}$ is sufficient in the epithelium to achieve the same smooth muscle intensity. How *Fgf9* expression and localization *in vivo* is regulated remains elusive.

The matching of the vasculature with the airways ensures efficient gas exchange. Blood vessel formation must therefore be tightly controlled and defects in this process often result in life-threatening diseases. Likewise smooth muscles are considered a key player in asthma, contributing to airway constriction and inflammation (Lazaar, 2002). Also during development, proper regulation of smooth muscles is crucial (Kim and Vu, 2006; Unbekandt et al., 2008). Our model highlights the potential of four proteins to form a core network that directs both lung branching, blood vessel formation and smooth muscle differentiation. A number of regulatory interactions remain to be established in experiments like those that regulate FGF9. In the absence of experimental data, we had to assume that *Fgf9* is expressed constitutively in the mesothelium which may not be true. It will be important to integrate future experimental results into the model to gain a more detailed understanding into this important regulatory process.

**Table 1. Values of dimensionless parameters used for simulations.**

| | Parameter | Mesothelium (1) | Mesenchyme (2) | Mesenchyme (3) | Epithelium (4) | Cavities (5) |
|---|---|---|---|---|---|---|
| diffusion | $D_S$ | 5 | 5 | 5 | 5 | 40 |
| | $D_{F10}$* | 1 | 1 | 1 | 1 | 40 |
| | $D_R$ | 0.02 | 0.02 | 0.02 | 0.02 | - |
| | $D_{F9}$ | 1 | 1 | 1 | 1 | 40 |
| | $D_C$ | 0.0001*** | 0.0001*** | 0.0001*** | 0.0001*** | - |
| | $D_V$ | 1 | 1 | 1 | 1 | 40 |
| complex | $\delta_C$ | - | 1.6 | 1.6 | 1.6 | - |
| | $v$ | - | 5 | 5 | - | - |
| degradation | $\delta_S$ | 0.2 | 0.2 | 0.2 | 0.2 | 0.2 |
| | $\delta_{F10}$ | 5 | 5 | 5 | 5 | 5 |
| | $\delta_R$ | 1 | 1 | 1 | 1 | - |
| | $\delta_{F9}$ | - | 0.005 | 0.005 | 0.005 | 0.005 |
| | $\delta_P$ | - | 0.1\|0.001** | 0.1\|0.001** | - | - |
| | $\delta_V$ | 5 | 5 | 5 | 5 | 5 |
| production | $\rho_S$ | - | - | - | 1300 | - |
| | $\rho_{F10}$ | - | 3.5 | 3.5 | - | - |
| | $\rho_R$ | - | 0.6 | 0.6 | - | - |
| | $\rho_{F9}$ | 100 | - | - | - | - |
| | $\rho_P$ | - | 1\|-** | 1\|-** | - | - |
| | $\eta$ | - | - | 0.5 | - | - |
| | $\rho_{V1}$ | - | 0.3 | 0.3 | - | - |
| | $\rho_{V2}$ | - | 1.5 | 1.5 | - | - |
| | $\rho_{V3}$ | - | 1.5 | 1.5 | - | - |
| regulation | $\kappa_{F9}$ | - | 2 | 2 | - | - |
| | $K_M$ | - | - | 0.1 | - | - |
| | $K_{F9}$ | - | - | 20 | - | - |
| | $K_{V1}$ | - | 2 | 2 | - | - |
| | $K_{V2}$ | - | 2 | 2 | - | - |

Domain parameters: $l_{ep} = 0.2$, $r_0 = 1$, $r_1 = 2$, $h_0 = 1$. $n = 2$, $m = 2$.

*Note that the diffusion coefficient of FGF10 $\bar{D}_{F10}$ in epithelium and inner radius of mesenchyme $r_0$ are used to nondimensionalize model.

**When two values are given, the first one is the value used in the tip of the bud; the second one is used in the stalk.

***Diffusion coefficient of cells (progenitors and smooth muscles) is set to very small values for computational purposes.

## Acknowledgements

The authors are grateful to Xin Sun and Eric Domyan for discussions. Denis Menshykau acknowledges ETH Zurich for an ETH fellowship.

## Competing Interests

The authors declare that there are no competing interests.